\documentclass[epj]{webofc}
\usepackage[varg]{txfonts}   % Web of Conferences font
\usepackage{url} 
\usepackage{subfig, lineno}
\newcommand{\ns}{$\rm \nu\,'\! s\ $}

\newcommand{\anse}{$\rm \bar{\nu}\,'\! s$}

\begin{document}
%
% \linenumbers
%
\title{Neutrino-nucleon cross sections at energies of Megaton-scale
detectors}

\author{A.
Gazizov\inst{1,2}\fnsep\thanks{\email{askhat.gazizov@desy.de}}
\and M. Kowalski\inst{1}
%\fnsep\thanks{\email{marek.kowalski@desy.de}}
\and K. S. Kuzmin\inst{3,4}
%\fnsep\thanks{\email{kkuzmin@theor.jinr.ru}}
\and V. A. Naumov\inst{3}
%\fnsep\thanks{\email{vnaumov@theor.jinr.ru}}
\and Ch.\ Spiering\inst{1}
%\fnsep\thanks{\email{christian.spiering@desy.de}}
}

\institute{%
DESY-Zeuthen, Platanenallee 6, D-15738 Zeuthen, Germany
 \and
Gran Sasso Science Institute, viale Francesco Crispi, 7, 67100
L'Aquila, Italy
 \and
Joint Institute for Nuclear Research, RU-141980, Dubna, Moscow region,
Russia
\and
Institute for Theoretical and Experimental Physics, RU-117218, Moscow,
Russia
}

\abstract{%
An updated set of (anti)neutrino-nucleon charged and neutral current 
cross sections at $3~{\rm GeV} \lesssim E_\nu \lesssim 100~{\rm GeV}$
is presented. These cross sections are of particular interest for the
detector optimization and data processing and interpretation in the
future Megaton-scale experiments like PINGU, ORCA, and Hyper-Kamiokande.
Finite masses of charged leptons and target mass corrections in
exclusive and deep inelastic $(\bar\nu)\nu N$ interactions are taken into
account. A new set of QCD NNLO parton density functions, the ABMP15,  
is used for calculation of the DIS cross sections. The sensitivity of 
the cross sections to phenomenological parameters and to extrapolations 
of the nucleon structure functions to small $x$ and $Q^2$ is studied. 
An agreement within the uncertainties of our calculations with 
experimental data is demonstrated. 
}
\maketitle
%%%%%%%%%%%%%%%%  Introduction   %%%%%%%%%%%%%%%%%
\section*{Introduction}
\label{intro}
Future Megaton-scale neutrino detectors, such as 
PINGU, % \cite{TheIceCube-PINGU:2014oha},
ORCA, % \cite{Katz:2014tta}, 
and Hyper-Kamiokande %\cite{DiLodovico:2015kta} 
can be used for determination of the neutrino mass hierarchy (NMH)
with the atmospheric neutrino fluxes. Depending on the NMH, the 
$\nu$-oscillations produce a few percent differences in the event
rate, appearing in a few-GeV energy region, since the matter effects 
act oppositely for \ns and \anse. The NMH signature relies on the fact
that both fluxes and cross sections are different for \ns and \anse, 
so that the combined event rates show the remaining NMH dependence.
For the proper interpretation of the data, the accurate predictions of
the atmospheric neutrino fluxes have to be paired with the reliable knowledge
of the (anti)neutrino-nucleon cross sections. We focus on the latter with an emphasis
on the energy range $E_\nu \gtrsim 3$~GeV and present a self-consistent 
set of theoretical models and phenomenological parameters allowing us to describe
the major contributions into the total ${\nu}N$ and ${\bar\nu}N$ cross sections at these energies. 
The results of the present study  may be included in future upgrade of the MC generator
ANIS~\cite{Gazizov:2004va} in order to extend its validity to lower energies 
and improve the predictions for high and ultrahigh energies.

%%%%%%%%%%%%%%  Models  %%%%%%%%%%%%%%%%%%%%%%%
\section*{Models and parameters}
\label{models}
All $\nu N$-interactions, both proceeding via charged current 
(CC) and neutral current (NC),
\begin{equation} 
\label{nuscat}
(\bar\nu_\ell)\nu_\ell + N \to   (\ell^+)\ell^-  +  X \quad {\rm (CC)}
\quad{\rm and}\quad
(\bar\nu_\ell)\nu_\ell + N \to (\bar\nu_\ell)\nu_\ell + X
\quad{\rm (NC)}
\end{equation}
($\ell=e,\mu,\tau$; $N=p,n$) may be classified according to
the number of mesons, pions, kaons, \emph{etc.}, appearing in the
final hadron state $X$. The total cross sections are combinations of 
the contributions from the channels \emph{(i)} with no pions, -- the 
(quasi)elastic scattering (ES or QES), \emph{(ii)} with one pion, -- 
the resonance single pion production (RES), and \emph{(iii)} inclusive 
or deep inelastic scattering (DIS) with $>2$ hadrons in the final 
state $X$: $\sigma_{\nu N}^{\rm tot} = \sigma_{ \nu N}^{\rm (Q)ES}
\oplus \sigma_{\nu N}^{1\pi}  \oplus \sigma_{\nu N}^{\rm DIS}$
(see, e.g., Ref.~\cite{Kuzmin:2006dt}).

For calculations of the (Q)ES contributions 
we use the standard approach with the nucleon axial mass parameter 
$M_A^{\rm QES}=1.02$~GeV extracted from available $\nu_\mu {\rm D}$, 
$\bar\nu_\mu {\rm H}$, and $\pi^\pm$ electroproduction data (for 
details, see Refs.~\cite{Bernard:2001rs,Bodek:2007ym,Kuzmin:2007kr} 
and references therein).

For obtaining of the RES contributions the extended Rein--Seghal
model \cite{Kuzmin:2003ji,Kuzmin:2004ya,Rein:2006di} is adopted  
with account for the pion-pole contribution to the hadronic axial 
current~\cite{Berger:2007rq}. The value of the axial mass parameter  
$M_A^{\rm RES} = 1.12$~GeV has been derived in Ref.~\cite{Kuzmin:2006dh} 
by fitting to the data available at that time. All known nucleon 
resonances with the masses below $\approx 2$~GeV are included and the 
interference of their amplitudes is properly taken into account
according to Ref.~\cite{Rein:1980wg}.

For the DIS cross sections we use the approach of 
Ref.~\cite{Albright:1974ts} with all 5 structure functions (SFs) being 
taken into account. In the case of CC processes this allows to account 
for the finite lepton mass, especially important for $\nu_\tau$. The 
functions $F_{1,2,3}(x,Q^2)$ are available by OPENQCDRAD-2.0 code 
\cite{Alekhin:2012sg}; we use it with the new NNLO (both for light 
and charm production) parton distribution functions of 
ABMP15~\cite{Alekhin:2015cza}. Transitions from $F_{1,2,3}(x,Q^2)$ to 
target mass corrected SFs $F_i^{\rm TMC}(x,Q^2)$ are performed 
according to Ref.~\cite{Kretzer:2003iu}. Below we do not discuss the 
complicated problem of the nuclear effects in the DIS SFs and focus 
only on purely (anti)neutrino-nucleon reactions.

To avoid a double counting, the phase spaces of RES and DIS are to be 
separated according to the mass of the final hadron state in
Eq.~(\ref{nuscat}), $W_{\rm cut} = m_X \gtrsim m_N + 2 m_\pi$. 
We found that $W_{\rm cut}=1.4$~GeV provides a good compatibility 
of these contributions: a variation of $W_{\rm cut}$ around
$1.4$~GeV brings to comparatively small variations of the sum 
$\sigma^{\rm RES}(E)+\sigma^{\rm DIS}(E)$  at all energies.

At intermediate energies significant uncertainties in calculations
of  the DIS cross sections arise from necessary extrapolations of 
$F_i^{\rm TMC}(x,Q^2)$ to small $Q^2$ where the perturbative QCD 
fails. In accordance with suggestion of Ref.~\cite{Capella:1994cr},  
we smoothly switch all SFs off for a given $x$ as $Q^2 \to 0$, 
assuming a power-law dependence on $Q^2$ (or on $\log(Q^2)$ in the 
case of $F_3^{\rm TMC}(x,Q^2)$); $F_i^{\rm TMC}(x,Q^2) = F_i^{\rm TMC}
(x,Q^2_{\min})\times [{Q^2}/(Q^2+b)]^{\alpha(1-x)}$. The parameter 
values $Q_{\min}^2 = 1.2$~GeV$^2$, $\alpha=0.5$ and 
$b=0.645$~GeV$^2$ provide a reasonable agreement with the data.

\begin{figure}%
\centering
\includegraphics[width=0.485\linewidth]{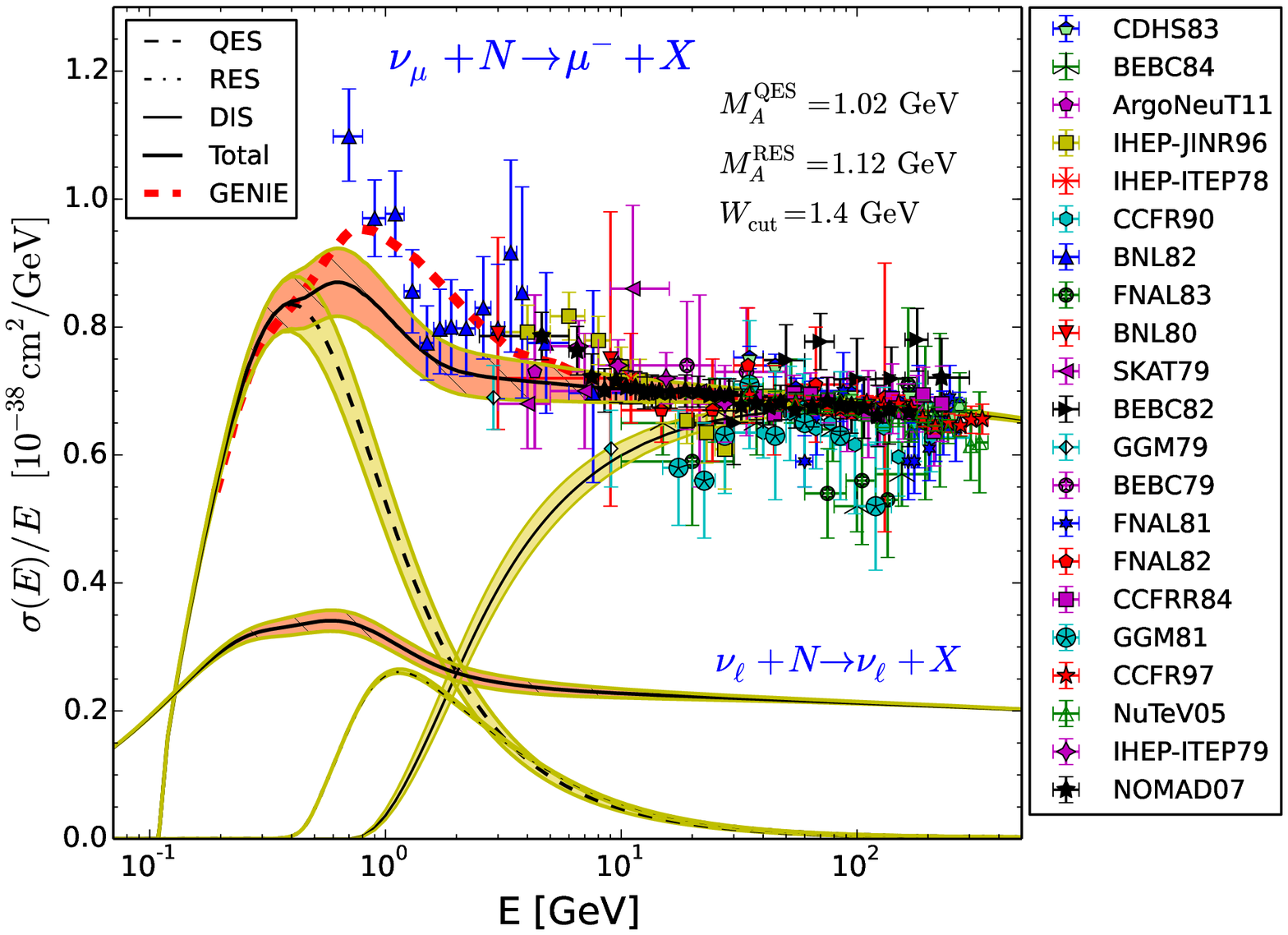}
\hfill
\includegraphics[width=0.485\linewidth]{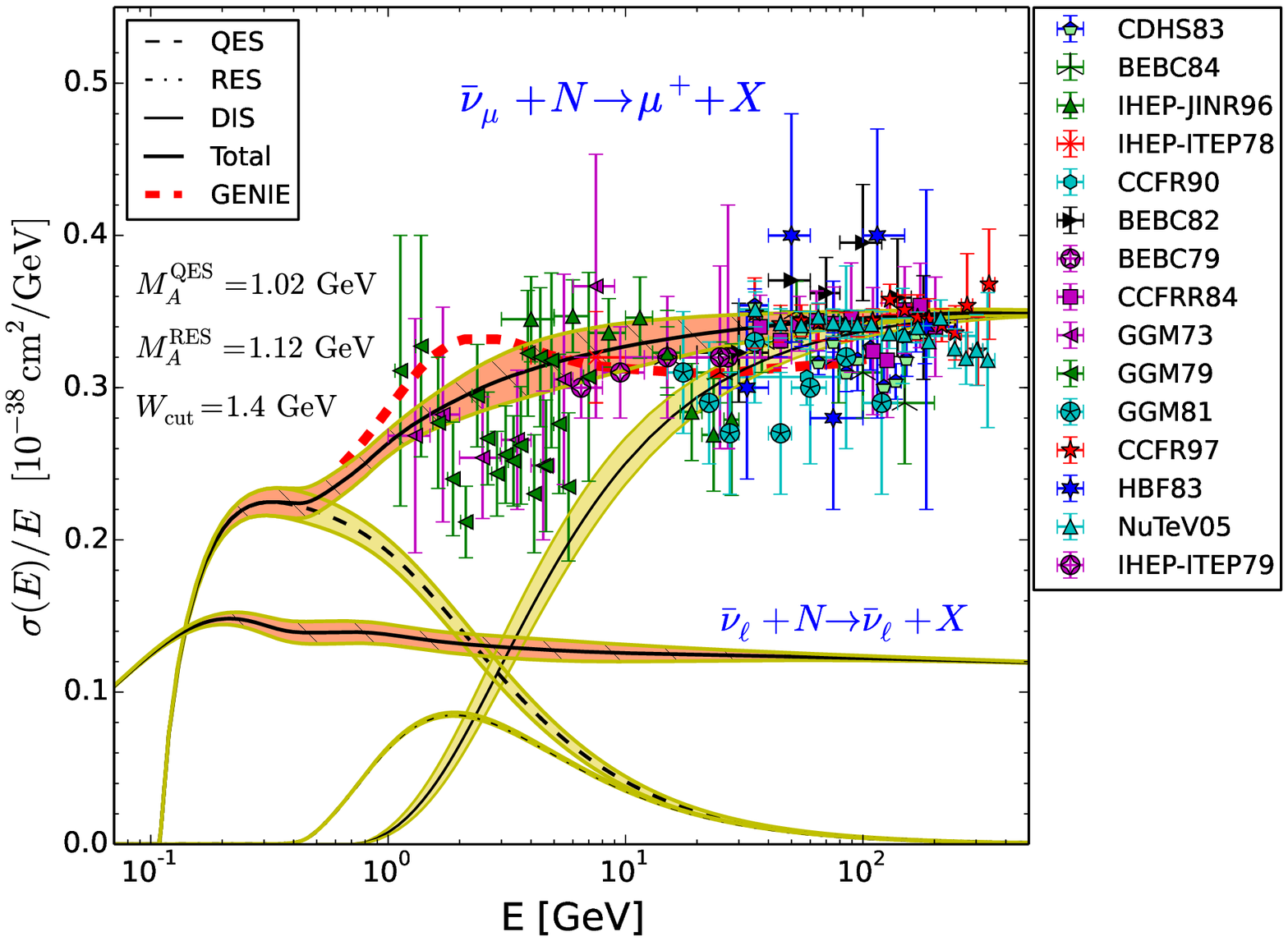} 
\caption{{\em Left panel:} Cross sections over energy for the CC 
$\nu_\mu N$ and NC scatterings off an isoscalar nucleon. Dashed, 
dash-dotted and thin solid lines correspond to contributions from QES, 
RES and DIS, respectively. Shaded bands show the uncertainties. Thick 
solid curve stands for the total cross section; the shaded hatched 
regions demonstrates the overall uncertainty. The thick dashed curve 
is the prediction of the GENIE neutrino MC generator 
\cite{Andreopoulos:2009rq}. {\em Right panel:} The same curves as in 
the left panel but for CC $\bar\nu_{\mu}N$ and NC scattering off the 
same target. The references to the data points can be found in 
Ref.~\cite{Kuzmin:2006dt}.    
}%
 \label{fig-1}%
\end{figure}

\begin{figure}%
\vspace*{-1mm}
\centering
\includegraphics[width=0.49\linewidth]{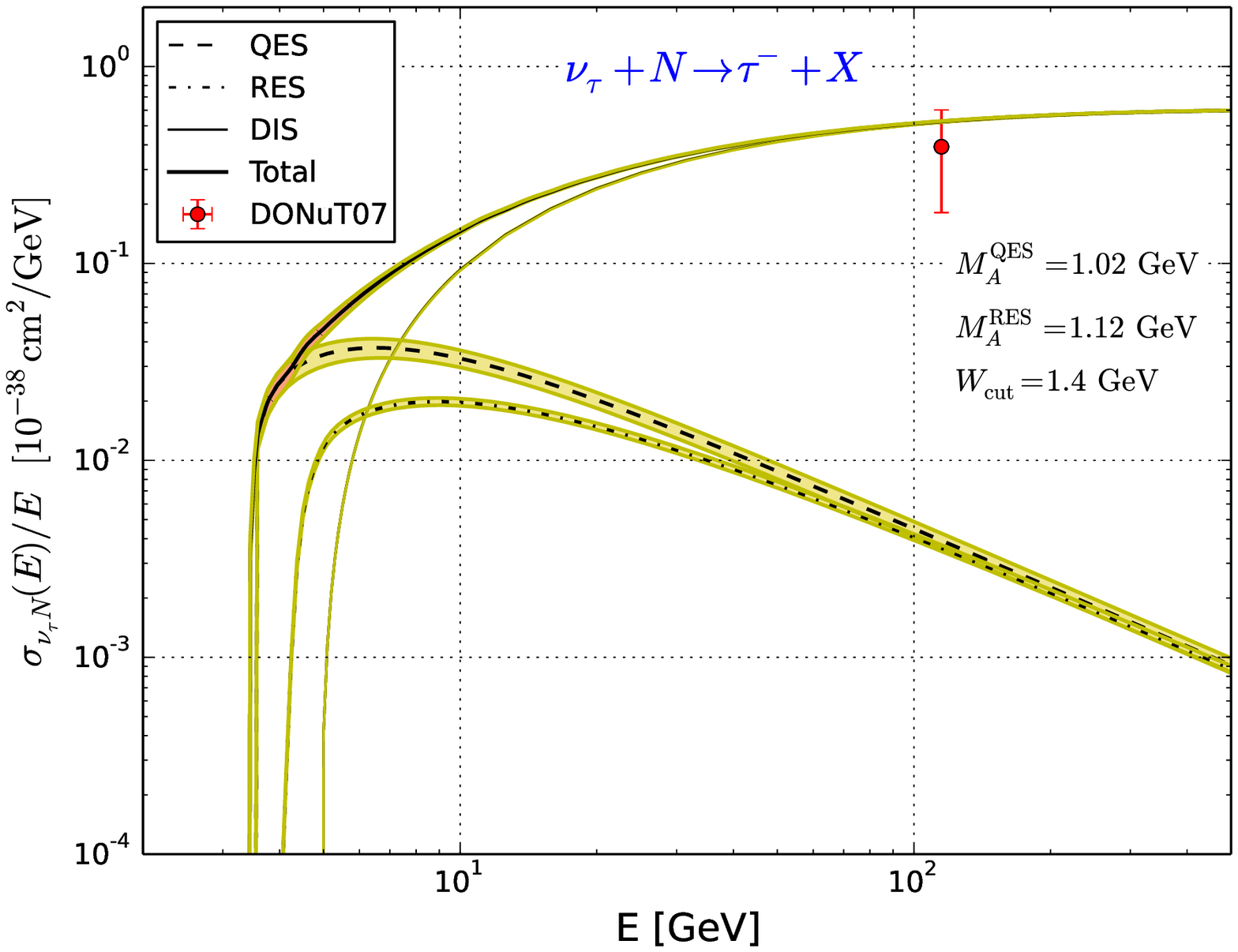}
\hfill
\includegraphics[width=0.49\linewidth]{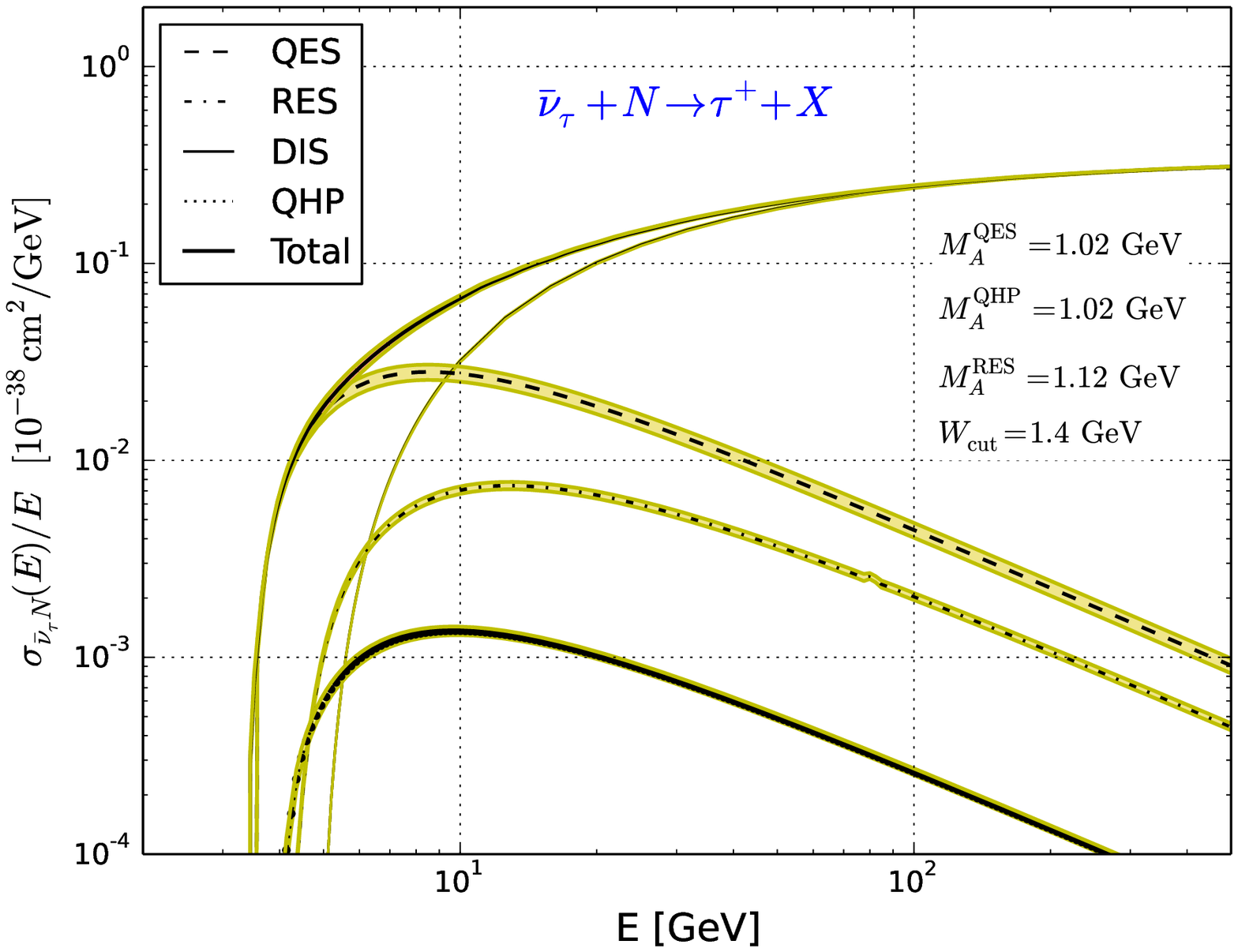}
\caption{The same curves as in the left panel of Fig.~\ref{fig-1} but 
for CC $\nu_{\tau}N$ and $\bar\nu_{\tau}N$ scatterings off an 
isoscalar nucleon. Additional dotted QHP line in the {\em right panel} 
shows a contribution from hypercharge violating QES production 
of lightest hyperons~\cite{Kuzmin:2008zz}. The DONUT data point is from 
Ref.~\cite{Maher:2005df}.
}%
\label{fig-2}%
\end{figure}

Cross section for $\nu_\mu$ and $\bar\nu_\mu$ scatterings off 
isoscalar nucleon calculated using our set of parameters are shown in 
Fig.~\ref{fig-1}. Analogous plots for the $\nu_{\tau}N$ and 
$\bar\nu_{\tau}N$ cross  sections are depicted in Fig.~\ref{fig-2}. 
All kinematic effects are taken into account. 
 
In Figs.~\ref{fig-1} and \ref{fig-2} the uncertainties for the QES and 
RES contributions are estimated by varying of the QES and RES axial 
mass parameters within the ranges $0.9~{\rm GeV} \leq M_A^{\rm QES} 
\leq 1.1~{\rm GeV}$ and $1.1~{\rm GeV} \leq M_A^{\rm RES} \leq  
1.3~{\rm GeV}$, respectively. The upper bounds of the bands in  
Fig.~\ref{fig-1} correspond to the higher values of the axial masses.
In order to estimate the DIS uncertainties, we vary the parameter 
$\alpha$ within the range $0.3$ to $0.7$. 


The solid curves in Fig.~\ref{fig-1} show similar predictions
of the Monte Carlo generator GENIE~\cite{Andreopoulos:2009rq}.
The reasons of our discrepancies with the GENIE predictions at low
energies are mainly in details of implementation of the Rein--Sehgal
model~\cite{Rein:1980wg} for the resonance single-pion 
neutrinoproduction and different cuts in $W$ used in the calculations.
This point will be discussed at length in a forthcoming paper.
Here we only mention the effect of finite lepton mass into the 
leptonic currents~\cite{Kuzmin:2003ji}, including the pion-pole 
contribution into the weak hadronic current~\cite{Berger:2007rq}, the 
effect of interference between the resonances having the same spin 
and orbital angular momentum of the final $N\pi$ state (both these 
effects are neglected in the cited version of GENIE), and also 
different ``recipes'' used for normalization of the Breit-Wigner 
(BW) factors and treatment of the unphysical BW ``tails''.
The disagreements with GENIE at high energies (most notable for 
$\bar\nu$) are due mainly to the PDF models involved in the GENIE 
code and in our calculations, and due to distinct methods for 
extrapolating the DIS SFs to small $Q^2$.

%%%%%%%%%%%%%  Conclusions %%%%%%%%%%%%%%%%%%%%
\section*{Conclusions}
\label{concl}
We presented a comprehensive set of the total cross sections for the 
neutrino and antineutrino scattering off protons and neutrons at 
energies most important for the future experiments with Mton-scale 
neutrino detector. Contributions from elastic and quasielastic 
scatterings (with ${\Delta}Y=0$ and $1$), single-pion 
neutrinoproduction, and deep-inelastic scattering were taken into 
account. For the ES, QES, and single-pion neutrino production 
contributions we rely on well-known methods and models. With the 
reasonable choice of parameters, our calculations are in reasonable 
agreement with the data within the experimental uncertainties. 
Calculations of the DIS cross sections were performed by using 
OPENQCDRAD-2.0 with the new set of NNLO PDFs of ABMP15 with account 
for the finite lepton masses and target mass corrections. We find 
that the uncertainties due the choice of the SFs extrapolations to 
small $Q^2$ are very significant. It is, in particular, responsible 
for the disagreements between our model and that used in the GENIE 
packet at high energies.

Present calculations will be used as a basis for the future upgrade 
of the Monte Carlo neutrino generator ANIS~\cite{Gazizov:2004va}.

%%%%%%%%%%%%   Acknowledgments %%%%%%%%%%%%%%%%%%%
\section*{Acknowledgments}
We are grateful to S.~Alekhin, J.~Bl\"{u}mlein and S.-O.~Moch for
making us available the new OPENQCDRAD and ABMP15 PDFs and to 
I.~Kakorin for a valuable contribution. A.G.\ thanks the Helmholtz 
Alliance for Astroparticle Physics, HAP for support. The work of 
K.K.\ and V.N.\ was partially supported by the Russian Foundation for 
Basic Research under Grant No.~14-22-03090.

%%%%%%%%%%%%%%%%%%%%%%%%%%%%%%%%%%%%%%%%%%%
%%
% \bibliography{nuncrs.bib}
% \end{document}

%%%%%%%%%%%    Bibliography   %%%%%%%%%%%%%%%%%%%%%
%%

%%%%%%%%%%%%%%%%%%%%%%%%%%%%%%%%%%%%%%%%%%%

\end{document}